\documentclass[mathleft,fleqn,%
]{an}
\usepackage{graphicx}
\usepackage[varg]{txfonts}
\usepackage[sort]{natbib}
\bibpunct{(}{)}{;}{a}{}{,}
\usepackage{tikz}
\def\tick{\tikz\fill[scale=0.4](0,.35) -- (.25,0) -- (1,.7) -- (.25,.15) -- cycle;} 
\newcommand{\kms}{km~s$^{-1}$}

\newcommand{\THe}{{$^3$He}}
\newcommand{\FHe}{{$^4$He}}

\begin{document}

% The following seven commands are intended for editorial usage and
% should be ignored by the author(s).
\Pagespan{1}{}% Document's page range. 
% If second parameter is left empty, the last page is computed
% automatically.
\Yearpublication{2016}%
\Yearsubmission{2015}%
\Month{0}%   
\Volume{999}%  
\Issue{0}% 
\DOI{asna.201400000}% 

\title{Observations of solar X-ray and EUV jets and their related phenomena}

\author{D.\, E. Innes\inst{1,2}\fnmsep\thanks{Corresponding author:
        {innes@mps.mpg.de}}
% Example for footnote, note the usage of the \texttt{fnmsep} command
% as separator between institute number and footnote mark}
\and R. Bu\u{c}\'i{k}\inst{1,3}
\and L.-J. Guo\inst{1,2}
\and N. Nitta\inst{4}
}
\titlerunning{Observations of solar jets}
\authorrunning{D.\,E. Innes}
\institute{
Max-Planck-Institut f\"ur Sonnensystemforschung, 37077 G\"ottingen, Germany
\and
Max Planck/Princeton Center for Plasma Physics, Princeton, NJ 08540, USA 
\and
Institut f\"ur Astrophysik, Georg-August-Universit\"at G\"ottingen, D-37077 G\"ottingen, Germany
\and
Lockheed Martin Advanced Technology Center, Palo Alto, CA 94304, USA}
\
\received{XXXX}
\accepted{XXXX}
\publonline{XXXX}


\abstract{%
 Solar jets are fast-moving, elongated  brightenings related to ejections seen in both images and spectra on all scales from barely visible chromospheric jets to coronal  jets extending up to a few solar radii. The largest, most powerful jets are the source of type III radio bursts, energetic electrons and ions with greatly enhanced \THe\ and heavy element abundances. The frequent coronal jets from polar and equatorial coronal holes may contribute to the solar wind. The primary  acceleration mechanism for all jets is believed to be release of magnetic stress via reconnection; however the energy  buildup depends on the jets' source environment. In this review, we discuss how certain features of X-ray and EUV jets, such as their repetition rate and association with radio emission,  depends on their underlying photospheric field configurations (active regions, polar and equatorial coronal holes, and quiet Sun).}

\maketitle

\section{Introduction}

Jets are transient, collimated and fast with respect to the thermal speed of the plasma in the jet. They are seen with innumerable different forms, scales and spectral signatures on the Sun. Here we focus on  X-ray \citep{Setal92, SSH98, Cirtain07} and extreme ultraviolet (EUV) jets \citep{AF99, Nistico09} and their associations with cold plasma surges \citep{CQWG99}, filaments \citep{Sterling15}, and solar energetic particles (SEPs) \citep{WPM06,Nitta15}. The consensus is that all solar jets result from magnetic reconnection and that the jet energy and characteristics depend on the magnetic environment and energy buildup process at the jet source. The aim of this review is to 
provide a  overview of the different types of jets from an observational point-of-view.
For example, what is the difference between jets caused by the motion of sunspot satellite flux \citep{Cetal96}, flux emergence \citep{Cheung15}, and those resulting from the winding up of fields surrounding minority flux in coronal holes  \citep{Guo13b, Schmieder13}? How do these compare with eruptions in the quiet Sun \citep{Innes09} or along coronal hole boundaries \citep{Madjarska04}?

 There are many types of jet and the boundaries between different categories are not  well-defined. Here we separate jets into two main types: interplanetary and coronal. The interplanetary jets are associated with sub-relativistic electron beams detected by decametric-hectometric (D-H) type III radio bursts and, depending on the magnetic connectivity of the source, SEPs. These are strong jets associated with a sufficiently high flux of energetic particles for them to produce a type III radio burst and they come from open field regions so that the accelerated particles can propagate into interplanetary space. 
 Those jets seen extending into the corona in X-ray or EUV images but without producing D-H type III's are categorised as coronal. Either they are strong jets but SEPs are ejected along closed field so return to the Sun, 
% Depending on the height of the loop, . 
 or they are not powerful enough to produce a sufficient flux of sub-relativistic electrons to create a type III.  

% If  SEPs are ejected along closed field, they will return to the Sun and, depending on the height of the loop, they may produce U-shaped radio bursts that reverse  frequencies above 100~MHz \citep{Karlicky96}. Less powerful jets may be seen as fast, collimated ejections in X-ray and EUV images which, if they are projected against the limb, may extend into white-light coronagraph images. 

There are many discussions of jets in the literature. The first surveys of X-ray jets was based on Yohkoh  observations  \citep{Shimojo96, Tsuneta91, SSH98}. Here  the jets were sorted according to the underlying magnetic field. Most jets (72\%) came from mixed polarity regions near pores or sunspots and 20\% came from unipolar or plage  regions near active regions with and without minor polarity flux concentrations.

Originally  X-ray jets were thought to be caused by flux emergence and subsequent reconnection with open magnetic flux \citep{Setal92}. Later it was observed that 
X-ray jets  may be associated with moving magnetic features (MMFs) and  co-temporal and spatial with spinning H$\alpha$\ surges \citep{Cetal96}, suggesting that  reconnection between the flux constraining twisted H$\alpha$ filament plasma, rather than emerging flux, and  adjacent open flux was the cause of the jets.  One of the questions addressed by our survey is how long after flux emergence are EUV and X-ray jets seen.  Many papers report a significant delay between jet formation  and flux emergence in contrast to recent 3D simulations of flux emergence into open field regions that predict  jets at the time of flux emergence \citep{Moreno08, Moreno13}.

Early observations of X-ray jets associated with  bursts of metric radio emission implied that the jet mechanism was able to accelerate sub-relativistic electrons along dense corona structures \citep{Aurass94,Kundu95,Raulin96}. 
Further confirmation came from observations of jets which were co-incident with hard X-rays from their footpoints \citep{Nitta97, Nitta08, Krucker11, Chen13} and from the jet itself \citep{Bain09,Glesener12}.

With the launch of Hinode \citep{Kosugi07} fainter X-ray jets in  polar and equatorial coronal hole regions could be investigated \citep{Cirtain07,Subramanian08,Raouafi10}.  
 \citet{Torok09} proposed that  coronal hole jets occur when
a small concentration of minor polarity flux emerges in a coronal hole creating an anemone-shaped system of loops centered on the minor polarity. When the system  is twisted about the central axis, stress builds up that can be explosively released, driving high-speed jets.

Another agent is believed to cause jets along the boundaries of large coronal holes. The open flux, rooted in the coronal holes, extends far into the corona where it rigidly co-rotates. But the underlying photospheric regions differentially rotate. The different speeds force the flux at coronal hole boundaries to continuously readjust and reconnect \citep{Wang93, Kahler02}, producing jets \citep{Madjarska04}.  
%The data from SDO are now providing valuable insights on jets and their triggering mechanism. 

Jets have also been seen from junctions of supergranular cells where vortices entrain mixed polarity flux creating conditions for small eruptions and jets \citep{Innes09,Innes13,Attie16}. 

Table 1 categorises jets recently reported in the literature. One criterion for inclusion in the table is that the jet was observed after 1994 when WIND/WAVES \citep{Bougeret95} data are available so that they can be identified as interplanetary (IP) or coronal (C). We also chose those jets with co-temporal and spatial photospheric magnetic field data, and for completeness a few publications reporting features of polar or limb jets.
 Column 1 gives the jet type and in brackets the source environment: SS - sunspot; MMF - moving magnetic features on the outer edge of sunspot penumbra;
 Plage -  plage; ARCH - active region coronal hole; PCH - polar coronal hole; ECH - equatorial coronal hole with quiet Sun field strengths; QS - quiet Sun.  Column 2 gives the type of SEPs, if any, associated with the jet. 
 All \THe\ jets were also associated with $40-100$~keV electrons and $0.4-1$~MeV~nucleon$^{-1}$ ions. Column 3 gives the time in hours between the main phase of flux emergence and the jet.  A '0' means that jets are produced during the main flux-emergence phase. Column 4 notes (with a tick) if flux cancellation below the jets was observed. Column 5 gives the time in hours between jets if more than 2 jets were observed from the same site. Columns 5, 6, and 7 note whether a surge, filament or circular ribbon was observed at the site of the jets. Blanks in the table indicate there was no data and an 'X' that there was data but the feature was not present. Where the interpretation of the data was ambiguous there is a question mark.
The tabulated features are chosen to distinguish the various scenarios discussed in the review.

\begin{table*}
\begin{center}	
	\begin{tabular}{l c c c c c c c c c}	
	\hline
	Jet type  		& SEPs			& Emerge		&Cancel 		& Multiple	&Surge	& Filament	&Circular 		& Reference	\\
				&				&  (hrs)		&			&(hrs)	&		&			&ribbon		&		\\
	\hline
	IP (SS)		& \THe\			&			&			& 6-12	&		&			& 			& \citet{Nitta08} \\
	IP (SS)		& 				&X			&			& 0.3-0.5	&X		&X			& X			& \citet{Innes11} \\
\hline
	IP (MMF) 		& 	 			&			&\tick		& 0.5-1.0	&\tick	&			&			&\citet{Chifor08}	  \\
	IP (MMF)		&				& 			&\tick		& 		&\tick	&			&			&\citet{Chen13}		\\
	IP (MMF)		& 				&			&			&		&\tick	&\tick		&			&\citet{LiX15} \\
	IP (MMF) 		& 				&1-2			&\tick		& 0.5-3.5	&\tick	&			&			&\citet{Chandra15}	  	\\
	IP (MMF) 		& 				&			&\tick		& 0.5-2.0	&\tick	&			&			&This work - 2012 July 02	  	\\	
\hline
	IP (Plage)		&\THe\			&36			&			& 		&\tick	&\tick		&\tick		&\citet{Buick15}		\\
         IP (Plage)		&				&24-48		&?			&		&\tick	&			&			&\citet{Schmieder13}\\
\hline
%	IP (ARCH)	& 				&			&			& 		&		&			&			&\citet{Kundu95}	\\
	IP (ARCH)	& \THe\			&24			&			& 		&		&			&			&\citet{Wang06}		\\
	IP (ARCH)	&				&48			&\tick		& 		&\tick	&\tick		&			&\citet{Liu11}		\\
	IP (ARCH)	&				&			&			&		&X		&X			&			&\citet{Alissandrakis15}\\
	IP (ARCH)	&\THe\			&36			&			& 18		&\tick	&\tick		&\tick		&\citet{Buick15}		\\
	IP (ARCH)	&\THe\			&60			&			& 		&\tick	&\tick		&\tick		&This work - 2014 May 16		\\
\hline
	C (MMF)		&				& 0		&			& 1-2	&\tick	&			&			&\citet{Cheung15}	\\
	C (MMF)		&				&X			&\tick		& 1-2	&\tick	&			&			&\citet{Chen15}		\\			
\hline
	C (Plage) 		&				&24-48		&?			&0.5-0.7	&\tick	&			&			&\citet{Guo13b}\\
\hline
	C (ARCH)		&				&			&			&0.5-1	&\tick	&			&\tick		&\citet{Zhang14}		\\
\hline
	C (PCH) 		&				& 			&			&		&\tick	&			&			&\citet{Cirtain07}	\\
	C (PCH) 		&				& 			&			&		&\tick	&			&			&\citet{Patsourakos08}	\\
	C (PCH) 		&				& 			&			&		&\tick	&			&			&\citet{Nistico09}	\\
%	CH			&				&			&			&		&		&			&			&				\\
	C (PCH) 		&				& 			&			&		&\tick	&			&			&\citet{Moore10}	\\
	C (PCH) 		&				& 			&\tick		&		&\tick	&			&			&\citet{Young14b}	\\
	C (PCH) 		&				& 			&			&		&\tick	&\tick		&			&\citet{Sterling15}	\\
\hline
	C (ECH)		&				& 3			& \tick		&$>1$	&		&			&			&\citet{Yang11} \\
	C (ECH)		&				&	6		&\tick		&		&		&			&\tick		&\citet{Huang12a}	\\
	C (ECH)		&				&			&			&		&\tick	&			&\tick		&\citet{Chandrashekhar14}	\\
\hline
        C (QS)		&				&			&\tick		&X		&\tick	&\tick		&			&\citet{Hong11}\\
	C (QS)		&				& X			&\tick		&X		&\tick	&			& 			&\citet{Innes10} \\
	C (QS)		&				& X			&\tick		& X		&\tick	&\tick		&\tick		& \citet{Adams14}	\\
	\hline
	\end{tabular}
	\caption{Characteristics of X-ray/EUV jets.  A description of the column entries is given in the text at the end of section 1.}
\label{table1}
\end{center}	
\end{table*}

\section{Interplanetary Jets}
The signature of interplanetary jets is a D-H (frequency 10-0.1 MHz) radio type III burst. Type III bursts are produced when sub-relativistic electron beams propagate from the Sun through interplanetary space. They produce bright bursts of radio emission that rapidly drift from above 10 to 0.1~MHz as a function of time (see, \citet{Reid14} and references therein). The drift speed depends on the exciter beam and the density decrease in surrounding plasma. The sub-relativistic, 0.02c-0.35c, electron beam \citep{Dulk87, Krupar15} excites Langmuir waves with a frequency proportional to the square root of the electron density. The radio brightness depends on both the energy of the electrons and their flux \citep{Dulk98}. Interplanetary jets  originate from open flux regions in the vicinity of active regions \citep{ Klein08}.  As well as 2-120~keV electrons \citep{Reames85, Benz01, Krucker99, Klassen11}, interplanetary jets also accelerate ions  (0.3 to 4~MeV~amu$^{-1}$). These ions are characterised by dramatically enhanced  (up to $10^4$) 
\THe/\FHe\ abundances \citep{Reames86, Pick06, WPM06, Nitta06} and moderately enhanced (up to 100) heavy ion abundances compared with the standard solar wind. 
Within active regions, interplanetary jets have been seen from the edge of sunspot umbra, the edge of the penumbra, plage and/or small active region coronal holes.  
%Aurass94, Kundu95,
%Recently attention has been to untwisting jets \citep{Pariat09, Lee13, Cheung15} but only 4\% of the X-ray jets were categorised as untwisting. Evidence for untwisting in active region jets is mostly from  spectra of the colder transition region plasma \citep{Cetal96, CQWG99, Lee13, Cheung15}.   
%One question is how the  jet properties depend on the  jet location within the active region.
\subsection{Sunspot}
\begin{figure}
\includegraphics[width=0.8\linewidth]{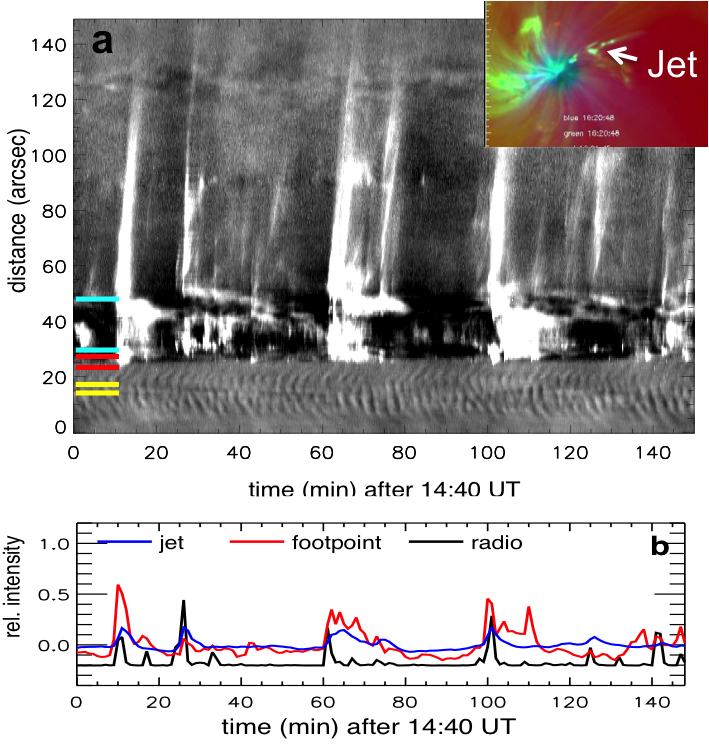}
%\includegraphics[width=\linewidth]{nitta08_ssjet.pdf}
%\caption{Jet on the edge of a sunspot. (a) Hinode/XRT soft X-ray reverse colour image. The white contour outlines the sunspot um- bra and penumbra. (b) Mass spectrograms from ACE/ULEIS for ions with energies 0.4-10 MeV nucleon?1. Images from Nitta et al. (2008).}
\caption{Jets on the edge of a sunspot \citep{Innes11}. (a) Time series of AIA 211~\AA\ intensity along the jet. On the left the red and blue lines indicate the loop footpoint and loop regions and the yellow lines are centred on the sunspot umbra. The inset shows a three color composite of line-of-sight magnetic field (red), 304~\AA\ (blue) and 211~\AA\ (green) at the time of the jet at 16:20 UT (100~min) on 3 Aug 2010. (b) Light curves of 1-10~MHz radio (black), and 211~\AA\ footpoint (red) and jet (blue) emissions. }
%\label{nitta_ssjet}
\label{innes_ssjet}
\end{figure}

Series of jets have been observed on the edge of sunspot umbrae \citep{Nitta08, Innes11}. \citet{Innes11} made a detailed analysis of a series of six interplanetary jets seen in 2.5 hours (Fig.~\ref{innes_ssjet}) and concluded that the jets were not triggered by flux emergence because no flux emergence was visible and there were no cold plasma surges alongside the the hot jets. The trigger may have been related to sunspot waves \citep{Sych15, Chandra15} forcing reconnection between the sunspot open flux and  closed field connected to opposite polarity satellite flux. 
%The footpoint brightening is simultaneous with the radio and precedes the EUV jets but about 30~s.
\THe -rich $0.4-1$~MeV nucleon$^{-1}$ ions and $40-200$~keV electrons have  also been detected from a sunspot jet  \citep{Nitta08}.

%\THe -rich SEPs have also been detected from a sunspot jet illustrated in Fig.~\ref{nitta_ssjet} \citep{Nitta08}.
%As can be seen in Fig.~\ref{nitta_ssjet} the hard X-ray (15-30~keV) photons are from the edge of the umbra. The bright soft X-ray emission are bright loops and the jet is seen as a faint soft X-ray spike coming out of the loops.

\subsection{Moving magnetic features} 
Strong interplanetary jets are frequently seen from the region just beyond sunspot penumbrae or pores \citep{Chifor08} where minor polarity satellite or  moving magnetic feature (MMF) flux reconnects with concentrations of main polarity flux. 
The cause of MMFs on the edge of sunspots  is not fully understood \citep{Thomas02, Hagenaar05, Rempel15}. Analysis of sunspot simulations led \citet{Rempel15}  to suggest that they result from a magneto-convection process related to the strength of the canopy field beyond the penumbra.

\begin{figure}
\includegraphics[width=0.8\linewidth]{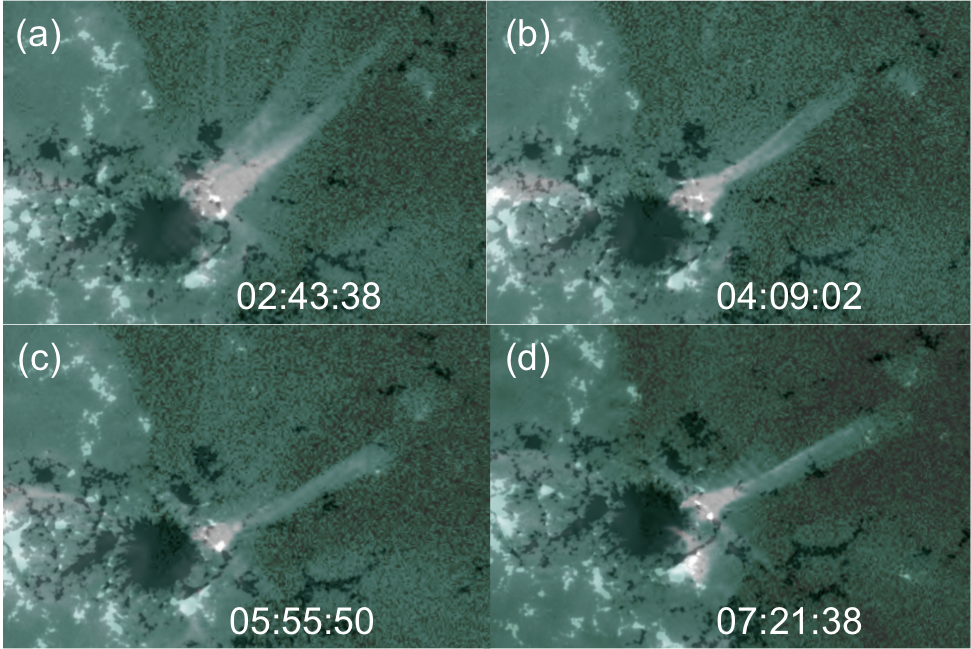}
\caption{SDO AIA 94~\AA\ images superimposed on the HMI longitudinal magnetic field for four interplanetary jets seen between 00:00 and 12:00 on 2012 July 02. A movie of the jets and MMFs is online.}
\label{4jets}
\end{figure}

\begin{figure}
\includegraphics[width=\linewidth]{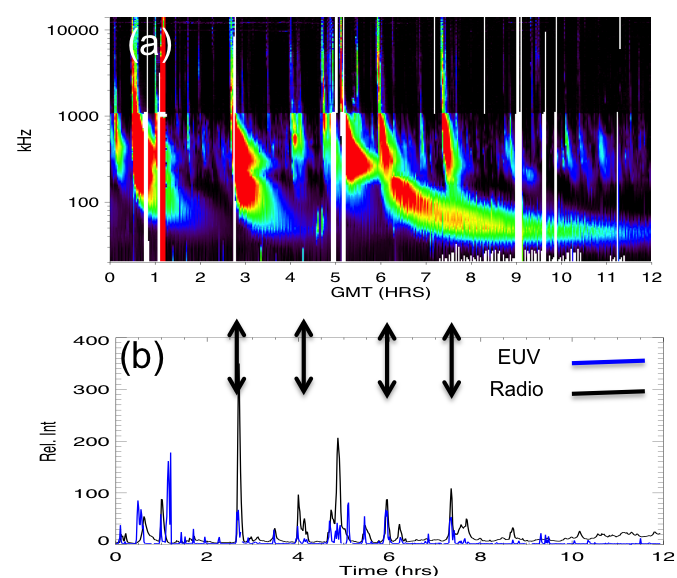}
\caption{Radio D-H type III bursts associated with jets from sunspot MMF  on 2012 July 02: (a) WIND/WAVES radio spectra; (b) light curve of 94~\AA\ emission (EUV) and integrated 1-10~MHz radio flux. The double headed arrows mark emission from jets shown in Fig.~\ref{4jets}.}
\label{wind}
\end{figure}

At the time of writing there were no recent good examples in the literature of jets from sunspot MMFs, but it is easy to find plenty of examples by browsing the SDO \citep{Lemen12, Scherrer12} data  with jhelioviewer  \footnote{See www.jhelioviewer.org}.  A typical example from 2012 July 02 is illustrated in Fig.~\ref{4jets}. Here the SDO/AIA  94~\AA\ images of four strong jets have been superimposed on the SDO/HMI longitudinal magnetic field. The WAVES radio data for the time period of the EUV jets are shown in Fig.~\ref{wind}(a). In the figure underneath  the 94~\AA\ light curve from the jet region is plotted together with the integrated 1-10~MHz radio emission.  There were seven interplanetary EUV jets in 7 hours. As can be seen in the online movie (120702\_jets), most jets do not occur when the flux emerges but from the slowly disappearing, and outward moving minor polarity (positive) fields. Once the minor polarity flux disappeared, the jets stopped. Later in the day, more positive flux appeared at the same site. Flux cancellation produced a series of coronal jets \citep{Chen15}. Although not shown all these jets were also associated with surges seen in 304~\AA.  \citet{Cetal96} described X-ray jets associated with H$\alpha$ surges overlying MMFs before the availability of WAVES data. It is possible that these were also interplanetary jets. 
The  cold plasma may be ejected as a result of flux emergence or it may be filament material that has formed during the evolution of the sunspot surroundings. 
We note that interplanetary jets from  MMFs and SSs are frequently seen in groups of more than three with an interval of 20~min to a couple of hours between jets implying a relatively short energy buildup time.

%There may be examples where flux emergence drives large interplanetary jets but the majority of jets occur when flux cancels, after shearing motions not during the flux emergence. 

\subsection{Unipolar plage and active region coronal holes}
The bright chromospheric emission close to and inside active regions is known as plage. The underlying magnetic fields are stronger than quiet Sun and often unipolar reflecting their origin which is either  dispersed  sunspot or  recently emerged active region field \citep{Borrero15}. 
They may develop into active region coronal holes (ARCHs) which are regions of open field with low coronal emission close to active regions. In both plage and ARCHs  magnetic fields concentrate along supergranular boundaries. 
%We found only one example of an interplanetary plage jet. It  had similar build up and features as several of the ARCH jets. 
Flux emergence is followed by at least 24 hours before a strong interplanetary jet is ejected. Post-jet loops often form a small circular patch of EUV loops with a broken ring of chromospheric ribbons at their footpoints. 

 Interplanetary jets from ARCHs may be recurring \citep{WPM06, Buick14} or single jets \citep{Liu11}.  Some are associated with filament eruption \citep{Liu11} but not all \citep{Alissandrakis15}. \citet{Liu11} suggest that flux emergence triggered the jet that they analysed  but inspection of the long-time evolution shows that flux emergence occurred 48~hours before the jet. Subsequently a filament formed along the neutral line between the emerged flux and coronal hole field. This was ejected with the jet.

\begin{figure}
\includegraphics[width=\linewidth]{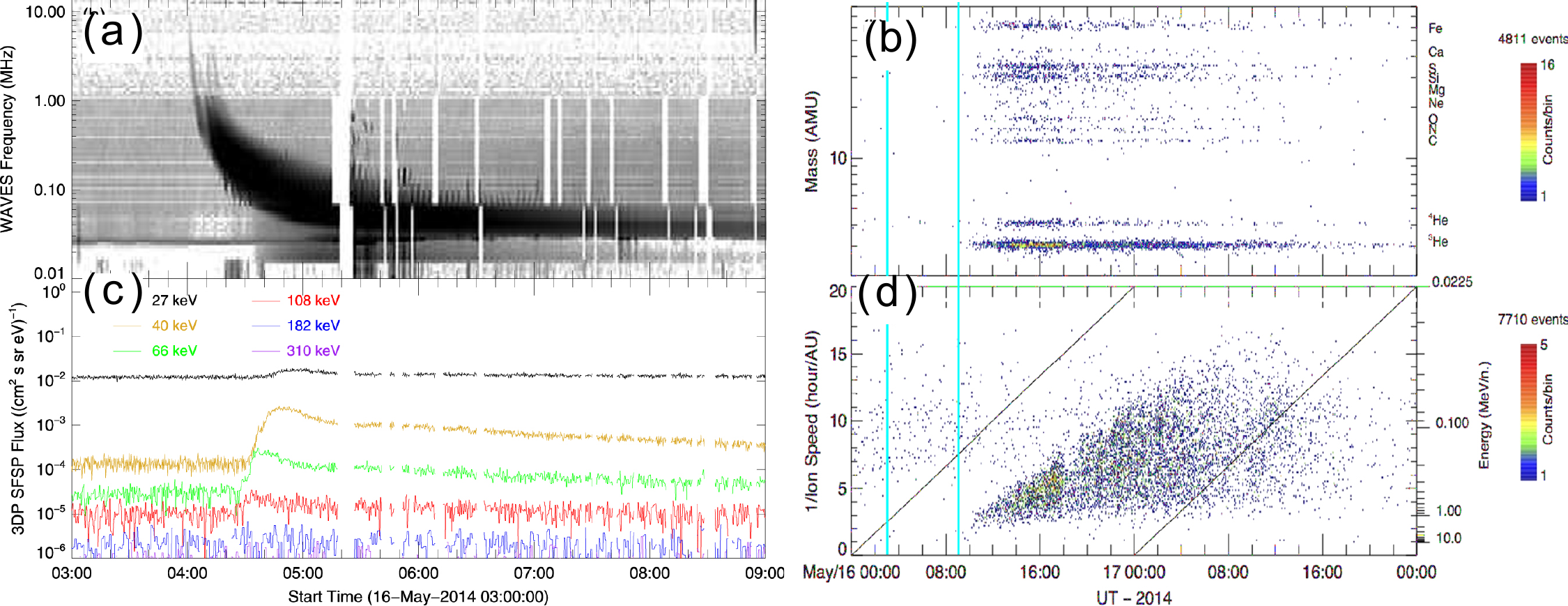}
\caption{ACE interplanetary signatures of jet on 2014 May 16 \citep{Nitta15}: (a) WAVES radio dynamic spectrum; (b) Mass spectrograms in energy range $0.4-10$~MeV~nucleon$^{-1}$; (c) 3DP Electron fluxes; (d) 1/ion speed for 
ions with mass between $10-70$~AMU. }
\label{seps_jet1}
\end{figure}

\begin{figure}
\includegraphics[width=\linewidth]{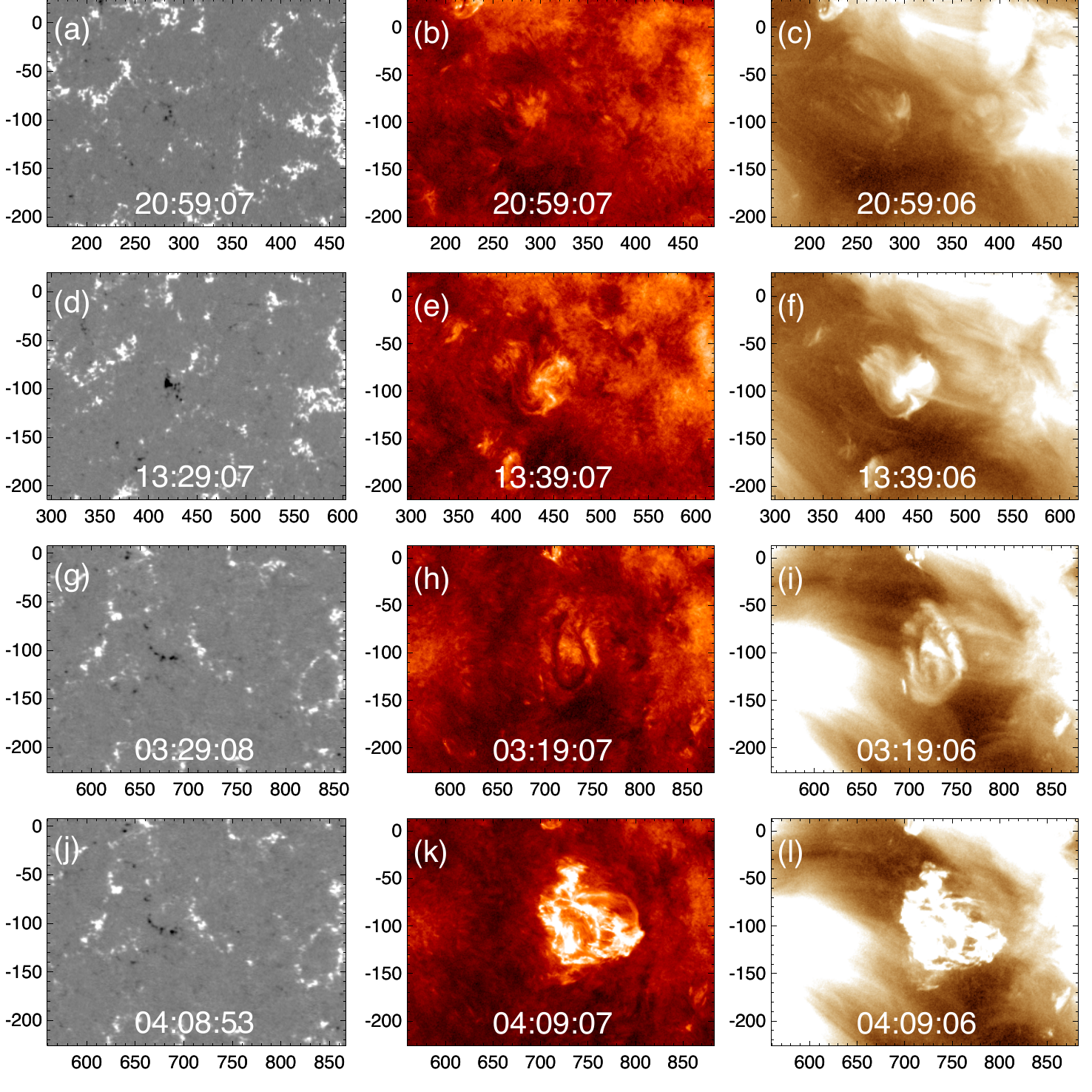}
\caption{Magnetic field (scaled between $\pm 250$~G) and EUV images of the ARCH jet on 2014 May 16: (left) magnetic field; 
(middle) AIA 304~\AA; (right) AIA 193~\AA. Dates are top row 2014 May 13, second row 2014 May 14, third and fourth rows 2014 May 16.}
\label{sdo_jet1}
\end{figure}

One of the plage jets and several of the ARCH jets were sources of  \THe-rich solar energetic particles. One  of the best examples, observed on 2014 May 16,  was reported by \citet{Nitta15}. It produced a \THe/\FHe\ enhancement of about $5\times10^4$ over that measured in the fast wind. The  interplanetary signatures (type III radio burst, SEPs with dispersive onset) of the jet, shown in Fig.~\ref{seps_jet1}, are typical of the other \THe\ jets in Table 1.
A few snapshots of the magnetic field, the chromospheric (304~\AA)  and coronal (193~\AA) plasma from the time of flux emergence to the jet are shown 
in  Fig.~\ref{sdo_jet1}. When the flux emerges, reconnection rapidly occurs between the minor polarity (negative in this case) and the surrounding major polarity flux, creating a series of radial loops connecting the minor polarity to the surrounding flux. A series of further small jets and brightenings were seen over a few days. Gradually a small, almost circular filament formed along the neutral line encircling the minor polarity. 
 Similar circular filaments were also  seen before flares producing circular H$\alpha$ ribbons \citep{Wang12}. The configuration is similar to the polar jet model of \citet{Pariat09} except here a filament forms along the neutral line. 
We note that during flux emergence no interplanetary jet was produced; however flux emergence inside the  unipolar region was necessary to  produce the embedded minor polarity and thus create the field structure suitable for interplanetary jet production. 

\begin{figure}
\includegraphics[width=\linewidth]{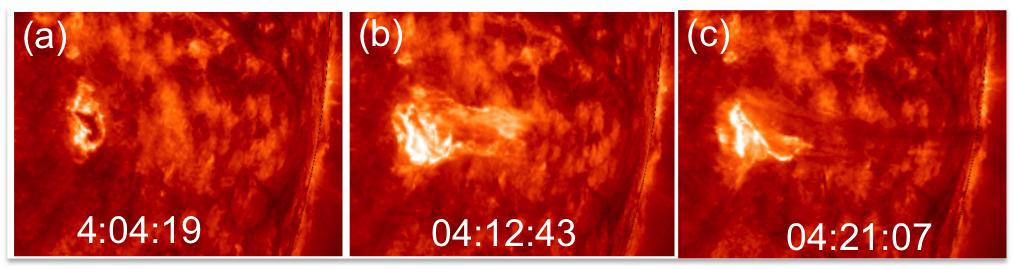}
\caption{EUV (304~\AA) images of the ARCH jet evolution on 2014 May 16: (left) pre-jet; 
(middle) onset; (right) helical structure.}
\label{untwist}
\end{figure}

The eruption occurred sequentially with filament sections lifting off one after another. If viewed from the correct angle and at the right time, helical structure in the cold plasma can be seen (Fig.~\ref{untwist}(c)).

In all studied plage and ARCH cases in Table 1, flux emergence was observed 1-3 days before the jet.
Unlike the SS and MMF jets, multiple jets are rare reflecting a longer buildup time for these types of jets. Filament eruption was seen in four cases. Since there was one case where no filament was seen, filament formation is not necessary for jet production. The filament may form  as a consequence of flux cancellation and shearing along the neutral line surrounding the minor polarity, as proposed by \citet{Balle89}. The filament then erupts at the time of the jet that started from the onset of an instability in the stressed configuration \citep{Pariat09}.
%Several equatorial coronal hole jets, especially near active regions show filaments eruption during the jet. 
%The \citet{Liu11} jet, although not reported in the paper also produced type III emission. 

%Recently \citet{Sterling15} have emphasised the importance of filaments in the initialisation of polar coronal hole jets on the limb. 

\section{Coronal Jets}
Coronal  jets  are seen to extend into the corona in X-ray and EUV images but do not produce D-H type III radio emission. Examples are X-ray jets from polar and equatorial coronal holes \citep{Cirtain07,  Subramanian08, Moore10} where the photospheric magnetic fields are typically quiet Sun values ($<100$~G). They also occur at the base of closed active region loops \citep{Strong92,Lee13,Cheung15}. Some produce metric U-type radio bursts \citep{Aurass94, Karlicky96} implying propagation of sub-relativistic electrons in  coronal loops. 

% lead to metric radio emissions \citep{Aurass94, Kundu95}.

\subsection{Active regions}
Many active region jets do not produce  interplanetary signatures either because they occur in closed-field regions or because they are too weak. 
For example \citet{Li12a} counted 575 jets from a recently emerged active region over 12 days but less than 10\% were associated with type IIIs. Like the interplanetary jets they generally have both hot and cold components which are not fully intermingled. Spectroscopic observations of the cold component sometimes show evidence for twist \citep{Cheung15}.  This may indicate recent flux emergence \citep{Cheung15}, the twisting of a fan-spine structure {\citep{Wyper15}, or  reconnection driven by MMFs. Not all active region coronal jets are twisted \citep{Chen16}. Future analysis of IRIS \citep{Depontieu14} and SDO data should be able to sort out the relationship between the hot and cold components and whether the jets appear with  flux emergence   and/or  energy and currents are built up in the corona by footpoint motions
\citep{Guo13b}.

\subsection{Coronal holes}
%At solar minimum  large well-defined coronal holes of opposite polarity are seen at the two poles. They form when poleward migrating flux from the active region belts  cancels with existing fields from the previous cycle. Sometimes, particularly during the decay phase of the cycle, equatorward  extensions are seen snaking down across the disk, generated by the dispersal and merging of active-region flux.
The polar coronal holes and their equatorward extensions are the source of many X-ray and EUV jets.
Coronal holes  rotate quasi-rigidly forced by the rigid rotation of the outer corona to which they are connected.  This causes  a buildup of current between the rigidly rotating open coronal flux and differentially rotating photospheric flux. It has been suggested that field lines continually reconnect leading to a high rate of EUV and X-ray jets  and brightenings along the boundary of coronal holes  \citep{Subramanian08, Yang11, Sako13}.  

%The cold jets, often called macrospicules, tend to have lower velocities, last longer and be shorter than the hot EUV and X-ray jets. 

\subsubsection{Polar}
X-ray and EUV jets from the polar coronal holes are often  projected against the corona which allows one to  see faint fast features and their continuation in white-light coronagraph images  \citep{WangYM98,Nistico09, Moore15}.
The plane-of-sky speeds of polar coronal hole X-ray jets ranges up to 800~\kms, the Alfv\'en speed in the lower corona, with an average speed of about 200~\kms \citep{Cirtain07}.
Because the EUV background is low, faint features at the jet origin can also be identified.  
 As with the active region jets, one frequently observes hot and cold components that appear to follow different paths \citep{Moore13}.
 Recently it has been suggested  by \citet{Sterling15} that  all polar jets start with  a filament eruption  because in all the edge-on jets that they studied in  polar coronal holes, a filament rises close to the base of the jet before footpoint brightening or jets in  X-ray and hot channel EUV emission are seen.
 As suggested in Section 2.3, the filament may be a  common bi-product of reconnection along a sheared neutral line that erupts with the jet, but it is  not necessarily vital for eruption. 
 
  %Filaments had been proposed previously  \citep{Raouafi10}. 

 % \subsubsection{Equatorial jets}
 
\citet{Pariat09,Pariat15} have suggested a model for coronal hole jets  that assumes  a minor polarity flux concentration in  a unipolar field. The minor polarity   connects to the dominant flux in all directions creating a circular fan of loops, as seen Fig.~\ref{sdo_jet1}(f) and (i). Sub-surface flows may then wind up the field, increasing its internal stress until it reaches a critical value and erupts. There is no filament in this model, so the eruption is caused by the opening of the coronal field not by a filament forcing the field opening.  It would be interesting to see if including a filament in the model would change the initiation process. 
%After eruption the process can repeat since the flux does not actually disappear. 

Support for the field opening scenario is provided by a unique series of STEREO \citep{Howard08} quadrature observations of a quiet Sun eruption taken with STEREO-A observing at high cadence (2.5~min) in 171~\AA\ and, simultaneously, STEREO-B in 304~\AA\ \citep{Innes10}. Thus response of the corona seen on the disk could be precisely compared to the cold-plasma eruptions on the limb (Figs.~\ref{dim_img} and \ref{dim_ts}).
Analysis of both quiet Sun and coronal hole events showed that the filament rise is nearly always preceded by dimming in the corona which  implies that the eruption starts with field opening and is followed by filament rise and footpoint brightening.

\begin{figure}
\includegraphics[width=0.8\linewidth]{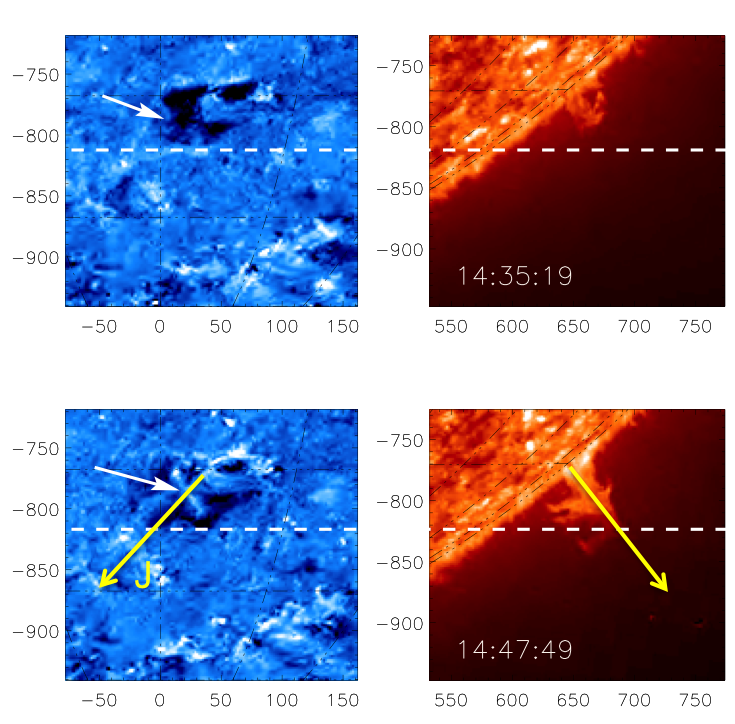}
\caption{STEREO-A 171 Å base difference, and simultaneous STEREO-B
304 Å intensity images of a quiet Sun eruption: onset (top row),
beginning of the fall back (bottom row). The yellow arrows in the bottom two images indicate the line along which the time series in Fig.~\ref{dim_ts} are taken.}
\label{dim_img}
\end{figure}

\begin{figure}
\includegraphics[width=0.8\linewidth, height=5 cm]{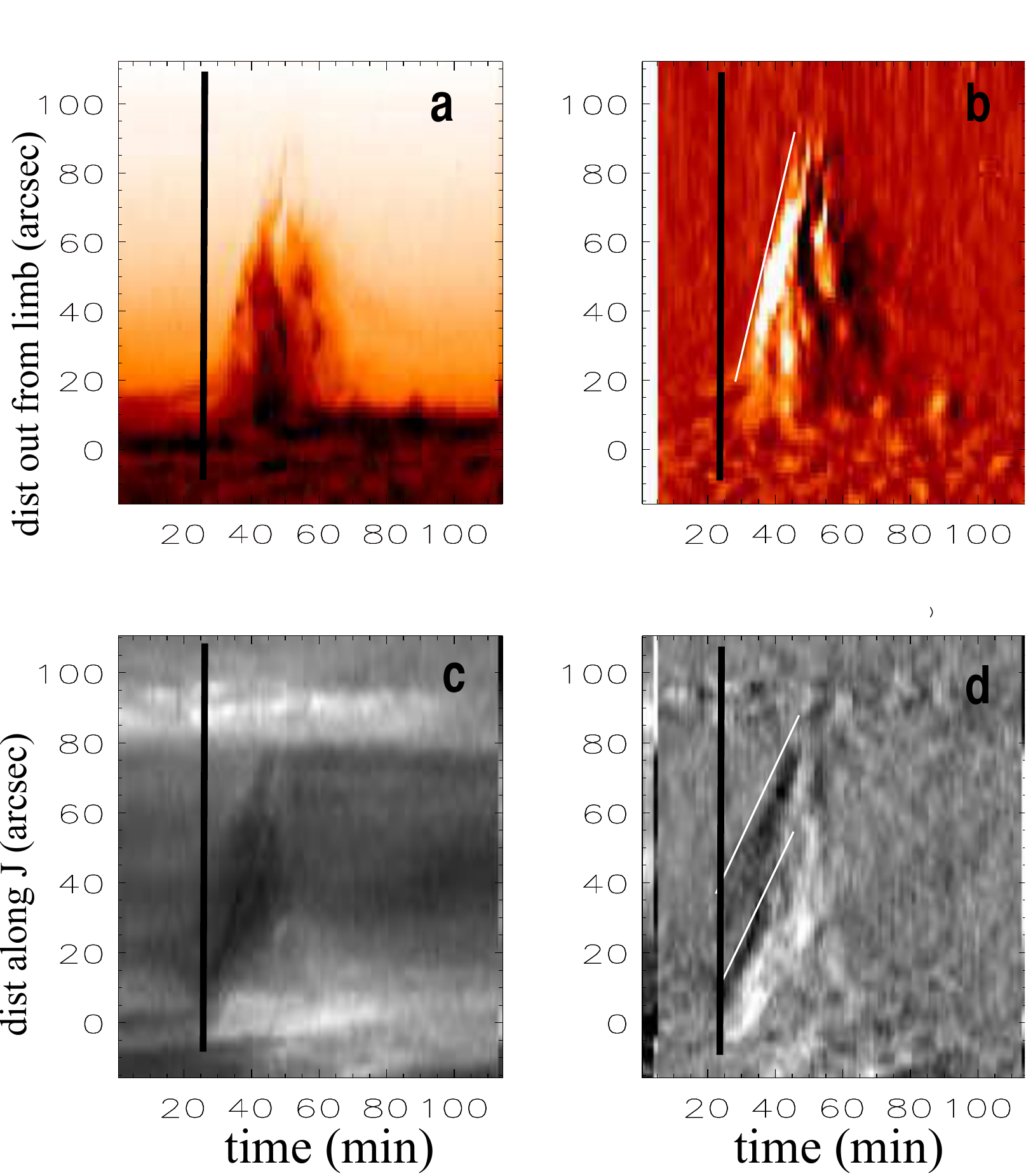}
\caption{ 304 Å and 171 Å time series. (a) and (b) STEREO-B 304 Å along the yellow arrow
in bottom right frame of Fig.~\ref{dim_img}: (a) intensity (negative); (b) running difference. (c) and (d) simultaneous
STEREO-A 171 Å along the yellow arrow in the bottom left image: (c) base difference;  (d)
running difference. The black vertical lines indicate the start time of the eruption seen in 171~\AA.}
\label{dim_ts}
\end{figure}

Other models, for example \citet{Moore10}, \citet{Raouafi10}, and \citet{Sterling15} propose a linear, arch-like, configuration with a loop connecting  the minority polarity and the coronal hole field in one direction, with a possibly sigmoid-shaped progenitor. This linear configuration would  be more likely if the original coronal hole flux is not uniformly distributed. Whether the reconnection occurs from arch-like or fan-like loops probably depends on the flux distribution in the plage/coronal hole surrounding the emerging flux. 
%Recently \citet{Sterling15} have highlighted the important role of filaments in the jet initiation. 
%If one wants to include filaments in the models, as observed by \citet{Sterling15}, then the filament should lie along the neutral line and be stabilised by a strapping field \citep{Low95}.  Indeed, as shown in Fig.~\ref{jet1} circular filaments do form around minor polarity flux concentrations. Also circular loops which are an   indication of fan-spine reconnection \citep{Pariat09} have been observed after plage, ARCH, ECH and QS jets.
Unfortunately, 
projection usually obscures, the magnetic field and chromospheric structure at the poles so it is difficult to verify if  the structure at the base of these jets is linear, as proposed by \citet{Moore13} or fan-like as proposed by \citet{Pariat09}.

Another common feature of large polar coronal hole jets is helical structure in the cold plasma (304~\AA) emission  \citep{Patsourakos08,Shen11,Moore13,Moore15}. 
The helical structure is explained as untwisting due to Alfv\'en waves generated during magnetic energy release  of an untwisting loop system or filament \citep{Cetal96, Pariat09, Raouafi10}.  An alternative explanation, demonstrated in Fig.~\ref{untwist}(c),   is  that sequential reconnection above a circular filament creates what looks like an untwisting structure.
%The observed twisted structure may be the result of delayed eruptions around a circular filament. 

\subsubsection{Equatorial}
Equatorial coronal holes allow the study of  the jets' magnetic environment and footpoint structure \citep{Huang12a,Chandrashekhar14}. 
%but the helical structure and plane-of-sky velocity of the jets themselves is more difficult  because the jets are projected against the disk. 
\citet{Huang12a} found that jets and brightenings are the result of magnetic flux cancellation about 6 hours after flux emergence.  
Brightening in and along the coronal hole boundaries is seen more frequently than in the quiet Sun \citep{Subramanian10} suggesting that either the open field structure is more conducive to reconnection, the brightenings are more visible due to lower background emission, or the flux emergence rate is different in the two regions. Understandably, many more of the brightenings (70\%) are jet-like compared to the quiet Sun (30\%). There are also
about 4-5 times more transition region jets along coronal hole boundaries than in the quiet Sun \citep{Madjarska04}. 

The large  number of obvious outflow events along the boundaries has reinforced the suggestion that coronal hole boundary outflows contribute significantly to the slow solar wind \citep{Wang98_wind,Woo04, Madjarska04}.

\section{Quiet Sun}
In the quiet Sun, supergranule flows cause entrained concentrations of magnetic flux to accumulate  at cell junctions. Here vortex-like flows may twist the field leading to current buildup and eruption \citep{Innes09}. As shown in Fig.~\ref{xray_qs} quiet Sun  X-ray brightenings and  small-scale eruptions occur  at the junctions of supergranular cells where vortices entwine opposite polarity flux concentrations \citep{Innes09,Innes13,Attie16}. Like the coronal hole jets reported by \citet{Sterling15} these eruptions also often have a filament at their core \citep{Adams14}. The jet-like nature of the eruptions has been revealed by spectroscopic observations \citep{Innes13}.

\begin{figure}
\includegraphics[width=\linewidth]{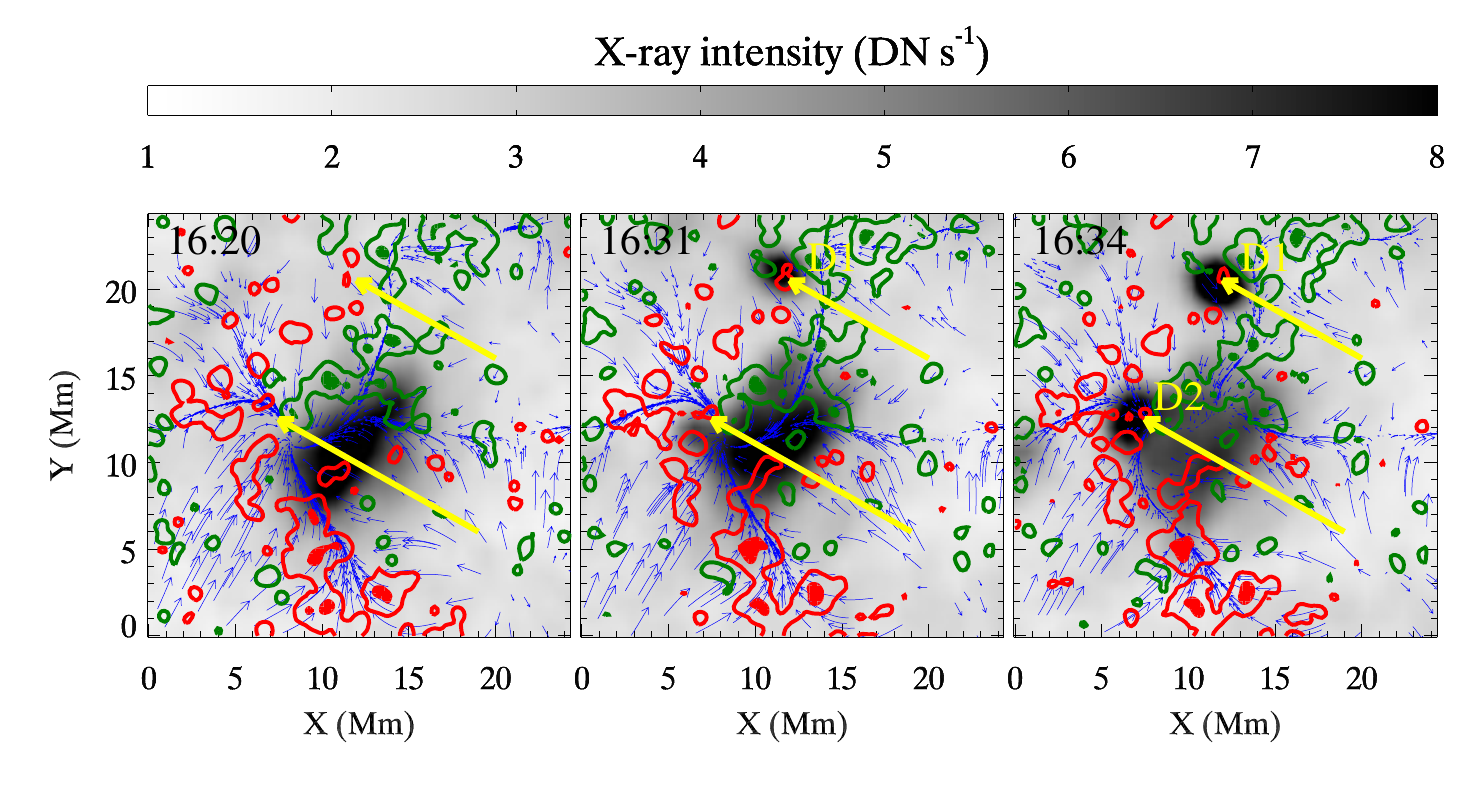}
\caption{ Snapshots of quiet Sun X-ray brightening \citep{Attie16}. X-ray images (gray colortable) and contours of  magnetograms. Red/green (respectively) is positive/negative polarity. Filled contours at +/-40 G, thin contours at +/-10 G. The blue arrows are the velocity vectors to show the direction of the flow. Their
length is scaled linearly with the magnitude of the flow. The yellow arrows point at the location of the X-ray transients.}
\label{xray_qs}
\end{figure}

\section{Summary}
We have categorised jets reported in the literature. Jets associated with sub-relativistic electrons capable of producing type III D-H radio bursts (interplanetary) have been observed  (i) on the edge of sunspot  umbra, (ii) just beyond the penumbra where there are opposite polarity moving magnetic features (MMF), (iii) from plage, and (iv) from active region coronal holes (ARCH). \THe-rich SEPs were detected from jets at the edge of sunspot umbra, plage and ARCH. So far there have been no specific reports of \THe\ from MMF jets but that is probably because MMFs are hard to see when the sunspots are at the nominal spacecraft connection point (about 55$^\circ$W). Because no polar coronal hole jet, not even the strongest observed, have produced type III bursts, it is likely that not enough sub-relativistic electrons are produced in polar coronal hole jets. We speculate that it is related to the weaker photospheric magnetic field strength.

Jets result from magnetic reconnection during (i) flux  emergence, (ii) on the edge of sunspot umbrae possibly forced by sunspot waves, (iii) between opposite polarity  MMFs, driven by sub-surface outflows from the sunspot, (iv) the winding up of spine-fan fields in unipolar flux regions, (v) at the junctions of supergranular cells where mixed polarity flux is entwined in vortex-like flows, and (vi) along the boarders of coronal holes forced by the different rotation speeds of the open and photospheric fluxes. Interplanetary jets from sunspots and MMFs have a significantly higher repetition rate than those from unipolar regions where a longer buildup phase seems to be required. 

Four types of trigger have been proposed: (i) flux emergence, (ii) filament rise, (iii) opening of coronal field  and, (iv) reconnection driven by footpoint motions. Most  observational analyses favour the filament  eruption. 
We found one recent paper where flux emergence was considered to be directly responsible for the observed jets and one that reported  evidence for coronal field opening.
Interestingly  recent simulations have concentrated on the flux emergence and coronal field opening rather than reconnection driven by footpoint motion.

 Future observations, particularly of the photospheric magnetic field  during the jet buildup phase, and  from more than one viewing  angle are needed to understand the triggering mechanism. Spectroscopic observations can give more complete information on flow motions and twist along jets. Also studies of SEP abundances compared to the source type and buildup time could lead to important information on the SEP acceleration process at reconnection sites.

\acknowledgements
 We thank the referee for the constructive comments. R.B. was supported by DFG grant BU 3115/2-1.
  
% Example of using BiBTeX (plus natbib):
% For details see \cite{1999MNRAS.309..731B},
% \cite{1893PASP....5..204C},
% \cite{2008IAUS..252...75L}. It has been demonstrated that this
% is important \citep{2012AN....333..663S}.

% Use this code if you wish to generate your bibliography with BibTeX;
% please replace first the string "an-demo" below with the name(s) of
% the BibTeX data base(s) you want to use.
% The resulting bibliography-output (the contents of the .bbl file)
% must be pasted into this file before submission.
% 
\bibliographystyle{an}
%\bibliography{/Users/innes/Documents/tex/sun2}
% 
% Replace the following example bibliography with your references
% before submission:

\end{document}